\documentclass[conference]{IEEEtran}
\IEEEoverridecommandlockouts

\usepackage{cite}
\usepackage{amsmath,amssymb,amsfonts}
\usepackage{graphicx}
\usepackage{textcomp}
\usepackage{xcolor}
\usepackage{xurl}
\usepackage{hyperref}

\usepackage{mathtools}

\usepackage{algorithm}
\usepackage{algpseudocode}

\usepackage{comment}
\def\BibTeX{{\rm B\kern-.05em{\sc i\kern-.025em b}\kern-.08em
    T\kern-.1667em\lower.7ex\hbox{E}\kern-.125emX}}

\usepackage[font=footnotesize,labelfont=bf,
            labelsep=space]{caption}

\DeclareMathOperator{\diag}{diag}

\usepackage[
    top=0.74in,
    bottom=1.05in,
    left=0.625in,
    right=0.625in
]{geometry}

\begin{document}

\title{Low-Complexity SVD-Based Precoding over Zero-Padded ISI Channels}

\author{\IEEEauthorblockN{Shuiyin  Liu}
\IEEEauthorblockA{\textit{Cyber Security Research and Innovation Centre} \\
\textit{Holmes Institute}\\
{Melbourne, VIC 3000, Australia} \\
Email: SLiu@Holmes.edu.au}
\and
\IEEEauthorblockN{ Amin Sakzad}
\IEEEauthorblockA{\textit{Department of Software Systems and Cybersecurity} \\
\textit{FIT, Monash University}\\
{Melbourne, VIC 3800, Australia} \\
Email: Amin.Sakzad@monash.edu}}

\maketitle

\begin{abstract}
We propose a low-complexity SVD-based transceiver for zero-padded memory-$1$ ISI channels. Appending a single zero-padding symbol to an $n$-symbol block yields an $n\times(n-1)$ channel matrix with an \emph{exact symmetric tridiagonal Toeplitz Gram matrix}. This structure provides closed-form singular values and vectors, enabling exact SVD precoding and equalization via FFTs without explicitly forming dense SVD matrices. The resulting implementation has $O(n\log n)$ complexity and incurs only a $1/n$ rate penalty. The resulting parallel scalar subchannels enable joint adaptive PAM and power allocation for SER minimization, followed by exact soft-information computation for LDPC decoding. At comparable code rates near $0.945$, simulations show that our proposed scheme outperforms a conventional TX-FFE/RX-DFE system with hard-decision Reed-Solomon (RS) decoding by approximately $5$ dB at FER $10^{-5}$. At high SNR, it achieves nearly the same FER performance as a SOVA-based receiver with LDPC decoding, while replacing trellis-based detection with fast SVD precoding and exact scalar soft-output computation.
\end{abstract}


\section{Introduction}

The rapid growth of AI and data-center traffic is driving Ethernet toward 800-Gb/s and $1.6$-Tb/s interfaces, with $400$-Gb/s-per-lane and even $448$-Gb/s-class links emerging as key targets \cite{ieee8023dj}. To achieve these rates within stringent bandwidth and power constraints, higher-order Pulse-Amplitude Modulation (PAM) signaling, including ongoing industry consideration of PAM-$6$ and PAM-$8$ for next-generation links \cite{graumann2025ieee}, is attracting increasing attention. However, higher-order PAM reduces the minimum constellation distance and increases sensitivity to channel loss, noise, and inter-symbol interference (ISI), creating a need for reliable ISI mitigation and low-complexity soft-decision Forward Error Correction (FEC) \cite{graumann2025ieee,Nebo2023,DFESOVA2024}.

Conventional transceiver architectures often rely on transmitter-side feed-forward equalization (TX-FFE) and receiver-side decision-feedback equalization (RX-DFE) to mitigate ISI \cite{DFEFFE2020}\cite{DFESOVA2024}. TX-FFE can effectively pre-compensate channel distortion, but under a fixed transmit-power constraint, part of the available power is used to generate ISI-canceling components rather than directly conveying information. For example, for a simple one-tap pre-equalizer \(x_i=s_i-\alpha s_{i-1}\) under average power limit \(\mathbb{E}[x_i^2]=P\), independent zero-mean symbols satisfy \(\mathbb{E}[s_i^2]=P/(1+\alpha^2)\), indicating that the energy available for $s_i$ is reduced by the pre-equalization operation. At the receiver, DFE can efficiently suppress post-cursor ISI using previously detected symbols, but erroneous decisions can propagate through the feedback loop and produce burst errors. Such bursty and correlated errors are particularly undesirable for soft-decision FEC, since they can distort the reliability information supplied to the decoder. Weighted decision-feedback equalization (WDFE) has also been investigated to exploit the reliability of previous hard decisions in the feedback loop \cite{WDFE2000}. However, the reliability weights generally require additional modeling of decision reliability. Overall, conventional TX-FFE/RX-DFE architectures offer low complexity but may sacrifice detection performanc, which can limit decoding performance for high-order PAM transmission.

Given these fundamental limitations, existing work has attempted to recover the lost performance by replacing/combining the simple DFE with more sophisticated detection and demodulation techniques, such as maximum-likelihood sequence detection (MLSD) \cite{Forney1972}, the soft-output Viterbi algorithm (SOVA) \cite{SOVA1989}, and other suboptimal soft-output detectors \cite{DFESOVA2024}, often in conjunction with powerful soft-decision FEC schemes such as LDPC. For an $M$-ary modulation over an ISI channel with memory $L$, trellis-based MLSD requires $M^L$ states and $M^{L+1}$ branches per symbol interval, resulting in a computational complexity that scales as $\mathcal{O}(nM^{L+1})$. For the widely studied memory-$1$ ISI channel ($L=1$), this reduces to $\mathcal{O}(nM^2)$. SOVA uses the same underlying trellis structure but incurs additional computational overhead for traceback and soft-output reliability updates. More recently, a reduced-complexity DFE-$3$ SOVA scheme employing a three-state trellis was proposed \cite{DFESOVA2024}. With traceback length \(\delta\geq 10\), its additional traceback complexity scales as \(O(nM_{\rm s}^2\delta)\), where \(M_{\rm s}\) denotes the number of trellis states. For the three-state DFE-3 implementation, this reduces to \(O(9n\delta)\), which can remain substantial for large \(\delta\). These limitations motivate the development of a fast receiver architecture with $\mathcal{O}(n\log n)$ complexity.

At a high level, TX-FFE/RX-DFE aims to suppress ISI and obtain more reliable scalar decisions, but typically use fixed PAM and equal power allocation. ML-based sequence detectors can approach optimal detection, but generally incur substantially higher computational complexity. Moreover, conventional schemes typically use fixed PAM order and equal power, without adapting to unequal channel gains. These observations motivate a fast transceiver architecture that exploits the exact channel structure while jointly adapting the PAM order and power allocation, achieving $\mathcal{O}(n\log n)$ complexity and $\mathcal{O}(n)$ memory usage, while delivering improved decoding performance over existing approaches.

In this work, we propose a fast SVD-based transceiver for zero-padded memory-$1$ ISI channels. By appending a single zero-padding symbol to each $n$-symbol transmission block, the resulting $n\times(n-1)$ channel matrix has a \emph{symmetric tridiagonal Toeplitz Gram matrix}, which admits closed-form singular values and singular vectors. This structure enables \emph{exact} SVD precoding through FFTs, without explicitly forming or storing dense SVD matrices. The resulting parallel scalar subchannels enable joint adaptive PAM and power allocation to minimize the worst-case symbol error rate (SER), together with exact soft-information computation for LDPC decoding. Overall, the proposed transceiver requires $\mathcal{O}(n\log n)$ complexity and $\mathcal{O}(n)$ memory, while incurring only a $1/n$ rate penalty due to zero padding. Table~\ref{tab:comparison} compares the complexity and performance of TX-FFE/RX-DFE, MLSD, SOVA, and the proposed SVD transceiver at a code rate near $0.945$ and $n\approx5440$.
\begin{table}[th]
\centering
\caption{Comparison of ISI Transceiver Schemes.}
\vspace{-2mm}
\label{tab:comparison}
\small
\setlength{\tabcolsep}{2.5pt}
\begin{tabular}{lccc}
\hline
Scheme & FEC & Complexity & 
$E_b/N_0$ @ FER $10^{-5}$\\
\hline
TX-FFE/RX-DFE \cite{DFEFFE2020} & RS 
& $\mathcal{O}(n)$ & $\simeq 19.8$ dB \\

RX-MLSD \cite{Forney1972} & RS 
& $\mathcal{O}(M^2n)$\textsuperscript{\(\dagger\)} & $\simeq 16$ dB \\

RX-SOVA \cite{SOVA1989} & LDPC 
& $\mathcal{O}(M^2n)$\textsuperscript{\(\dagger\)} & $\simeq 14.8$ dB \\

\textbf{Proposed SVD} & LDPC 
& $\mathcal{O}(n\log n)$ 
& $\simeq 14.5$ dB \\
\hline
\end{tabular}
\vspace{1mm}
\raggedright
\footnotesize
\textsuperscript{\(\dagger\)}: For the $M=8$ PAM memory-$1$ implementation used in the simulations.
\vspace{-6mm}
\end{table}

\section{System Model}

\subsection{Memory-1 ISI Channel}

We consider a real-valued memory-$1$ ISI channel over $n$ consecutive channel uses, given by
\begin{equation}
y_i=x_i+\alpha x_{i-1}+z_i,\qquad i=1,\ldots,n,
\quad x_0=0,
\label{eq:isi_model}
\end{equation}
where $\alpha$ denotes the ISI coefficient and
$z_i\sim\mathcal{N}(0,\sigma_z^2)$ is additive white Gaussian noise. The memory-$1$ model captures the effective ISI after transmitter-side pre-equalization and therefore provides a fundamental building block for longer-memory channels.
The channel-input symbols $x_i$ are generated from pulse-amplitude modulation (PAM) information
symbols through the precoding. Specifically, let
\begin{equation}
\mathbf{s}=[s_1,\ldots,s_{n}]^T,
\qquad s_i\in\mathcal{X}_{M_i},
\end{equation}
where $\mathcal{X}_{M_i}$ denotes the zero-mean, equally spaced
$M_i$-ary PAM alphabet with Gray labeling, i.e.,
\begin{equation}
\mathcal{X}_{M_i}
=
\left\{
-(M_i-1),-(M_i-3),\ldots,
M_i-3,M_i-1
\right\}.
\label{eq:M_i}
\end{equation}
Let \(\mathbf{p}=[p_1,\ldots,p_n]^T\) denote the PAM spacing coefficients and define \(u_i=p_i s_i\). Then, the scaled PAM vector is $\mathbf{u}=\mathbf{p}\odot\mathbf{s}=[u_1,\ldots,u_n]^T$, where \(\odot\) denotes element-wise multiplication, and the channel input is generated as
\begin{equation}
\mathbf{x}
=\operatorname{Enc}(\mathbf{p}\odot\mathbf{s})=\operatorname{Enc}(\mathbf{u}).
\label{eq:encoder}
\end{equation}

We impose the total transmission-power constraint
\begin{equation}
\mathbb{E}\!\left[|\mathbf{x}|^2\right]=nP,
\label{eq:power_constrain}
\end{equation}
where \(P := E_s\) denotes the average energy per channel use.

For a block of $n$ consecutive channel uses, the channel can be represented in vector form as
\begin{equation}
\mathbf{y}=\mathbf{H}\mathbf{x}+\mathbf{z}=\mathbf{H}\cdot \operatorname{Enc}(\mathbf{u})+\mathbf{z},
\label{eq:vector_isi_model}
\end{equation}
where $\mathbf{x},\mathbf{y},\mathbf{z}\in\mathbb{R}^{n}$ and the channel matrix is the lower-bidiagonal matrix
\begin{equation}
    \mathbf{H}=
    \begin{bmatrix}
        1 & 0 & \cdots & 0 \\
        \alpha & 1 & \ddots & \vdots \\
        0 & \alpha & \ddots & 0 \\
        \vdots & \ddots & \ddots & 1
    \end{bmatrix}.
    \label{eq:H}
\end{equation}

Thus, the ISI channel over a block of $n$ uses is a structured linear transformation, providing the basis for the low complexity precoding and equalization schemes below.

\subsection{TX-FFE/RX-DFE with Equal PAM and Power Allocation}
\label{sec:ffe_dfe}

\subsubsection{TX-FFE}
For the fixed-PAM baseline, we use a common PAM order $M_i=M$ and equal power allocation $p_i=p$ for all channel uses. At the transmitter, a $t$-tap FFE is represented by the lower-triangular Toeplitz matrix
\begin{equation}
    \mathbf{F}_t
    =
    \begin{bmatrix}
        f_0 & 0 & 0 & \cdots & 0 & 0 & \cdots & 0 \\
        f_1 & f_0 & 0 & \cdots & 0 & 0 & \cdots & 0 \\
        f_2 & f_1 & f_0 & \ddots & \vdots & \vdots & & \vdots \\
        \vdots & \ddots & \ddots & \ddots & 0 & \vdots & & \vdots \\
        f_{t} & \cdots & f_2 & f_1 & f_0 & 0 & \cdots & 0 \\
        0 & f_{t} & \cdots & f_2 & f_1 & f_0 & \ddots & \vdots \\
        \vdots & \ddots & \ddots & \ddots & \ddots & \ddots & \ddots & 0 \\
        0 & \cdots & 0 & f_{t} & \cdots & f_2 & f_1 & f_0
    \end{bmatrix}.
    \label{eq:F_t}
\end{equation}
Note that $t$-tap FFE means \(t\) ISI-canceling taps plus the main tap, i.e., \(t+1\) coefficients \(f_0,\ldots,f_t\). The FFE output is
\begin{equation}
\mathbf{x}=\beta_t\mathbf{F}_t\mathbf{u},
\label{eq:ffe_output}
\end{equation}
and the corresponding received vector is
\begin{equation}
\mathbf{y}
=\mathbf{H}\mathbf{x}+\mathbf{z}
=\beta_t\mathbf{H}\mathbf{F}_t\mathbf{u}+\mathbf{z},
\label{eq:effective_ffe}
\end{equation}
where $\beta_t$ is chosen to satisfy the total transmission-power constraint in \eqref{eq:power_constrain}. Specifically, $\beta_t=
{\sqrt{nP}}/
{\sqrt{\mathbb{E}\!\left[|\mathbf{F}_t\mathbf{u}|^2\right]}}$. To  cancel the ISI in \eqref{eq:isi_model}, the FFE coefficients are chosen as
\begin{equation}
f_k=(-\alpha)^k,
\qquad k=0,\ldots,t.
\end{equation}

The normalization therefore enforces the total average transmission-energy constraint. Let $\hat{\mathbf{H}}_t=\mathbf{H}\mathbf{F}_t$. For the memory-$1$ ISI channel, $\hat{\mathbf{H}}_t$ has the structure
\begin{equation}
    [\hat{\mathbf{H}}_t]_{i,j}
    =
    \begin{cases}
        1, & i=j,\\
        -(-\alpha)^{t+1}, & i=j+t+1,\\
        0, & \text{otherwise},
    \end{cases}
    \qquad i,j=1,\ldots,n.
    \label{eq:Hhat_entries}
\end{equation}
Hence, the $t$-tap FFE cancels the first $t$ ISI terms, leaving only the $(t+1)$-th-order residual ISI.

\subsubsection{RX-DFE}
At the receiver, the residual post-cursor ISI occurs at delay $t+1$, and a DFE can successively cancel it using previously detected symbols through successive interference cancellation (SIC). Specifically,
\begin{equation}
r_i
=
y_i
+
\beta_t(-\alpha)^{t+1}\hat{u}_{i-t-1},
\qquad
i=t+2,\ldots,n,
\end{equation}
where $\hat{u}_{i-t-1}$ denotes the corresponding previous hard decision. Assuming that all $i-1$ previous decisions are correct, the $i^{\text{th}}$ channel reduces to an AWGN channel,
\begin{equation}
r_i=\beta_t u_i+z_i.
\label{eq:FFEDFE_channel}
\end{equation}

\emph{Limitations of TX-FFE/RX-DFE:}
Although increasing the number of FFE taps, $t$,  progressively suppresses the ISI, it also reduces the effective signal amplitude due to the power normalization. In particular, $\beta_t<1$ for $t\geq1$ and decreases with $t$, while the noise variance remains unchanged. Consequently, the effective SNR of the channel in \eqref{eq:FFEDFE_channel} is reduced by a factor of $\beta_t^2$. Thus, conventional TX-FFE trades ISI suppression for an SNR loss. Moreover, the SIC-based DFE relies on previously detected symbols; an erroneous decision can therefore propagate to subsequent decisions, resulting in error propagation and degraded detection performance.

\subsection{ML-Type Detectors: MLSD and SOVA}

Instead of making hard decisions successively as in the RX-DFE,
maximum-likelihood sequence detection (MLSD) jointly detects the
transmitted PAM sequence by exploiting the residual ISI structure
\cite{Forney1972}. From the matrix representation in
\eqref{eq:effective_ffe}, the ML sequence is given by
\begin{equation}
\hat{\mathbf{x}}_{\mathrm{ML}}
=
\arg\min_{\mathbf{x}\in\mathcal{X}_M^n}
\left\|
\mathbf{y}
-
\beta_t\hat{\mathbf{H}}_t\mathbf{x}
\right\|^2,
\label{eq:MLSD}
\end{equation}
where $\mathcal{X}_M$ denotes the PAM-$M$ constellation. For a finite-memory channel, this optimization can be represented by a trellis, with each state determined by the previous transmitted symbols, and solved using the Viterbi algorithm (VA).

However, trellis-based ML detection has complexity that grows rapidly with both the channel memory and PAM order. For a memory-$L$ ISI channel with PAM-$M$ signaling, the trellis has $M^L$ states and $M$ outgoing branches per state, resulting in $\mathcal{O}(nM^{L+1})$ branch-metric computations and $\mathcal{O}(M^L)$ state-metric storage for the MLSD and SOVA. For the memory-$1$ channel considered here, this reduces to $\mathcal{O}(nM^2)$, i.e., quadratic growth with the PAM order.

\subsection{Problem Statement}

The preceding discussion motivates the design of a transceiver that
combines the reliability and soft-output capability of ML-type
detection with the low complexity of conventional equalization.
Specifically, we seek a finite-length ISI transceiver with
$\mathcal{O}(n\log n)$ computational complexity and
$\mathcal{O}(n)$ memory, while preserving soft information for
subsequent channel decoding and avoiding the rapidly increasing
complexity associated with trellis-based detection.

\section{Fast SVD-Based Equalization for Finite-Length ISI}
\label{sec:fast_svd}

We first introduce a single-zero-padding transformation that converts the finite-length ISI channel into a structured channel with a closed-form singular system. The resulting right singular vectors form a DST-I basis and can be implemented efficiently using FFTs. Hence, exact SVD-based precoding and equalization can be performed in $\mathcal{O}(n\log n)$ complexity and $\mathcal{O}(n)$ memory, without explicitly forming the singular vectors.

Zero padding has previously been combined with SVD-based transceivers \cite{SVD_ZP_Fading2007} and low-complexity pre-equalization for block transmission \cite{ZF_ZP2010}. In contrast, we exploit zero padding to obtain an effective ISI channel whose Gram matrix has a symmetric tridiagonal Toeplitz structure, enabling a closed-form SVD and its low-complexity FFT implementation.

\subsection{Single-Zero Padding Transformation}

Considering the memory-$1$ ISI channel in \eqref{eq:isi_model} and its matrix form in \eqref{eq:vector_isi_model}, we impose a single zero-padding constraint
\begin{equation}
x_n=0.
\label{eq:dst_termination}
\end{equation}
The transmitted vector contains $n-1$ symbols.
\begin{equation}
\widetilde{\mathbf{x}}
=
[x_1,x_2,\ldots,x_{n-1}]^{T},
\label{eq:dst_xtilde}
\end{equation}
while the physical transmitted vector is $\mathbf{x}=[\widetilde{\mathbf{x}} ^T,0]^T$.

The resulting channel can be written as
\begin{equation}
\mathbf{y}
=
\mathbf{H}\mathbf{x}+\mathbf{z}
=
\widetilde{\mathbf{H}}\widetilde{\mathbf{x}}+\mathbf{z},
\label{eq:dst_effective_channel}
\end{equation}
where $\widetilde{\mathbf{H}}\in\mathbb{R}^{n\times(n-1)}$ is
\begin{equation}
\widetilde{\mathbf{H}}
=
\begin{bmatrix}
1&0&0&\cdots&0\\
\alpha&1&0&\cdots&0\\
0&\alpha&1&\ddots&\vdots\\
\vdots&\ddots&\ddots&\ddots&0\\
0&\cdots&0&\alpha&1\\
0&\cdots&0&0&\alpha
\end{bmatrix}.
\label{eq:dst_A}
\end{equation}

The key structural property is that the channel Gram matrix $\mathbf{T}=\widetilde{\mathbf{H}}^{T}\widetilde{\mathbf{H}}
$ is symmetric tridiagonal Toeplitz:
\begin{equation}
\mathbf{T}
=
\begin{bmatrix}
1+\alpha^2&\alpha&0&\cdots&0\\
\alpha&1+\alpha^2&\alpha&\ddots&\vdots\\
0&\alpha&1+\alpha^2&\ddots&0\\
\vdots&\ddots&\ddots&\ddots&\alpha\\
0&\cdots&0&\alpha&1+\alpha^2
\end{bmatrix}.
\label{eq:dst_toeplitz}
\end{equation}

This structure permits a closed-form characterization of the singular system \cite{TTmatrices2013} and, consequently, fast SVD-based precoding and equalization. While the proposed single-zero padding introduces one redundant channel use per block, the resulting rate loss, $1/n$, vanishes as $n$ grows.

\subsection{Closed-Form Singular System of $\widetilde{\mathbf{H}}$}
Consider the compact SVD
\begin{equation}
\widetilde{\mathbf{H}}
=
\mathbf{U}
\mathbf{\Lambda}
\mathbf{V}^{T},
\label{eq:dst_svd}
\end{equation}
where $\mathbf{U}\in\mathbb{R}^{n\times(n-1)}$, $\mathbf{\Lambda}\in\mathbb{R}^{(n-1)\times(n-1)}$, and $\mathbf{V}\in\mathbb{R}^{(n-1)\times(n-1)}$. Let $\mathbf{\Lambda} =\diag (\lambda_1,\ldots,\lambda_{n-1})$, then
\begin{equation}
\mathbf{T}
=
\mathbf{V}
\mathbf{\Lambda}^{2}
\mathbf{V}^{T}.
\label{eq:dst_eigendecomposition}
\end{equation}

\subsubsection{Closed-Form $\mathbf{V}$}
The eigenvectors of $\mathbf{T}$ form the orthonormal Discrete Sine Transform of Type I (DST-I) basis \cite{TTmatrices2013}. Specifically,
\begin{equation}
[\mathbf{V}]_{j,k}
=
\sqrt{\frac{2}{n}}
\sin\left(
\frac{jk\pi}{n}
\right),
\qquad
j,k=1,\ldots,n-1.
\label{eq:dst_V}
\end{equation}
Since the DST-I matrix is symmetric,
\begin{equation}
\mathbf{V}^{T}
=
\mathbf{V}.
\label{eq:dst_V_symmetric}
\end{equation}

\subsubsection{Closed-Form $\mathbf{\Lambda}$}
The eigenvalues of $\mathbf{T}$ are \cite{TTmatrices2013}
\begin{equation}
\lambda_k^2
=
1+\alpha^2
+
2\alpha\cos\left(
\frac{k\pi}{n}
\right),
\qquad
k=1,\ldots,n-1,
\label{eq:dst_lambda_squared}
\end{equation}
and hence the singular values of $\widetilde{\mathbf{H}}$ are
\begin{equation}
\lambda_k
=
\sqrt{
1+\alpha^2
+
2\alpha\cos\left(
\frac{k\pi}{n}
\right)
},
\qquad
k=1,\ldots,n-1.
\label{eq:dst_lambda}
\end{equation}

\subsubsection{Closed-Form $\mathbf{U}$}
The left singular-vector matrix need not be explicitly constructed. It follows directly from
\begin{equation}
\mathbf{U}
=
\widetilde{\mathbf{H}}
\mathbf{V}
\mathbf{\Lambda}^{-1},
\label{eq:dst_U}
\end{equation}
and therefore
\begin{equation}
\mathbf{U}^{T}
=
\mathbf{\Lambda}^{-1}
\mathbf{V}^{T}
\widetilde{\mathbf{H}}^{T}.
\label{eq:dst_UT}
\end{equation}
Thus, both singular-vector transformations can be implemented without explicitly forming $\mathbf{U}$ or $\mathbf{V}$.

\subsection{FFT-Based Implementation of the DST-I Transform}

For any $\mathbf{q} \in \mathbb{R}^{n-1}$, the right singular-vector matrix $\mathbf{V}$ is the orthonormal
DST-I matrix, so both $\mathbf{Vq}$ and $\mathbf{V}^T \mathbf{q}$ can be
implemented efficiently using FFTs.

For computing the $k^{\text{th}}$ element in $\mathbf{Vq}$, the DST-I is
\begin{equation}
[\mathbf{V}{\mathbf{q}}]_k
=
\sqrt{\frac{2}{n}}
\sum_{j=1}^{n-1}
q_j
\sin\left(\frac{jk\pi}{n}\right),
\quad k=1,\ldots,n-1.
\label{eq:dst_matrix_vector}
\end{equation}
Define the odd-symmetric extension
\begin{equation}
\mathbf{q}_{\mathrm e}
=
\begin{bmatrix}
0&
q_1&\cdots&q_{n-1}&
0&
-q_{n-1}&\cdots&-q_1
\end{bmatrix}^{T}.
\label{eq:dst_extension}
\end{equation}
Under the standard FFT convention, if
$\mathbf{\Omega}=\operatorname{FFT}(\mathbf{q}_{\mathrm e})$, then
\begin{equation}
-\operatorname{Im}\{\Omega_{k+1}\}
=
2\sum_{j=1}^{n-1}
q_j\sin\left(\frac{jk\pi}{n}\right), \quad \text{hence},
\label{eq:dst_fft_identity}
\end{equation}
\begin{equation}
[\mathbf{V}{\mathbf{q}}]_k
=
-\sqrt{\frac{1}{2n}}\,
\operatorname{Im}\{\Omega_{k+1}\},
\qquad k=1,\ldots,n-1,
\label{eq:dst_fast}
\end{equation}
which implements the DST-I using a single length-$2n$
FFT. Since $\mathbf{V}^T=\mathbf{V}$, the same construction computes
$\mathbf{V}^T\mathbf{q}$ for any $\mathbf{q}\in\mathbb{R}^{n-1}$.
We denote this transform by
\begin{equation}
\mathbf{V}\mathbf{q}
=
\mathbf{V}^T\mathbf{q}
:=
\operatorname{DST-I}_n(\mathbf{q}).
\label{eq:dst_operator}
\end{equation}

\subsection{Fast SVD-Based Transceiver via FFT}

Let the scaled PAM symbols be
\begin{equation} \widetilde{\mathbf{u}} =[p_1s_1,\ldots,p_{n-1}s_{n-1}]^T, \qquad s_i\in\mathcal{X}_{M_i}. 
\end{equation}

\subsubsection{Transmitter}

The SVD precoder first applies the DST-I transform to the
information-bearing symbols:
\begin{equation}
\widetilde{\mathbf{x}}
=
\mathbf{V}\widetilde{\mathbf{u}}
=
\operatorname{DST-I}_n(\widetilde{\mathbf{u}}).
\label{eq:dst_precoding}
\end{equation}
A single zero is then appended to form the transmitted block
$
\mathbf{x}
=
[\widetilde{\mathbf{x}}^T,0]^T
$, consequently, 
\begin{equation}
\mathbf{y}
=
\mathbf{H}\mathbf{x}+\mathbf{z}
=
\widetilde{\mathbf{H}}\mathbf{V}\widetilde{\mathbf{u}}
+\mathbf{z}
=
\mathbf{U}\mathbf{\Lambda}\widetilde{\mathbf{u}}
+\mathbf{z}.
\label{eq:dst_diag_channel}
\end{equation}
Thus, the single-zero padding together with the DST-I precoder
realizes the exact SVD of the structured channel without explicitly
forming $\mathbf{V}$.

\subsubsection{Receiver}

The receiver first computes
\begin{equation}
\mathbf{q}
=
\widetilde{\mathbf{H}}^T\mathbf{y},
\qquad
q_j=y_j+\alpha y_{j+1},
\quad j=1,\ldots,n-1,
\label{eq:dst_rx_q}
\end{equation}
which requires only $\mathcal{O}(n)$ operations. The SVD projection
is then obtained using the same FFT-based DST-I transform:
\begin{equation}
\mathbf{r}
=
\mathbf{U}^T\mathbf{y}
=
\mathbf{\Lambda}^{-1}
\mathbf{V}^T\mathbf{q}
=
\mathbf{\Lambda}^{-1}
\operatorname{DST-I}_n(\mathbf{q}).
\label{eq:dst_rx}
\end{equation}
Since $\mathbf{U}$ has orthonormal columns, the projected noise
$\hat{\mathbf{z}}=\mathbf{U}^T\mathbf{z}$ remains white Gaussian
with the same variance. Hence, the $n-1$ SVD modes are exactly
decoupled into parallel AWGN channels,
\begin{equation}
r_i=\lambda_i p_is_i+\hat{z}_i,
\qquad i=1,\ldots,n-1.
\label{eq:dst_rx_s}
\end{equation}

\vspace{-2mm}
\subsection{Computation Complexity and Storage}
The singular values $\{\lambda_k\}_{k=1}^{n-1}$ are obtained directly from \eqref{eq:dst_lambda} in $\mathcal{O}(n)$ operations. The matrix-vector products involving $\mathbf{V}$ and $\mathbf{V}^{T}$ are implemented using the length-$2n$ FFT construction above, requiring $\mathcal{O}(n\log n)$ operations, while the multiplication by $\widetilde{\mathbf{H}}^{T}$ requires $\mathcal{O}(n)$ operations. Thus, the complete SVD-based precoding and equalization requires 
$$
\mathcal{O}(n\log n)
$$
computational complexity and $\mathcal{O}(n)$ working memory, since neither singular-vector matrix is explicitly formed.

\section{Adaptive PAM, Power Allocation, and LLR Computation}
\label{sec:adaptive_pam}

The SVD-based transceiver in Section~\ref{sec:fast_svd} decomposes the zero-padded ISI channel into $n-1$ parallel AWGN subchannels in \eqref{eq:dst_rx_s}. The resulting subchannels enable adaptive PAM and power allocation (APPA), followed by exact soft-information computation. Since the APPA parameters $\{M_i,p_i\}_{i=1}^{n-1}$ are computed once for a given noise variance $\sigma_z^2$ and reused across transmission blocks, the per-block transceiver requires $\mathcal{O}(n\log n)$ complexity and $\mathcal{O}(n)$ memory, dominated by the proposed FFT-based SVD.

\subsection{Adaptive PAM and Power Allocation (APPA)}
\label{sec:adaptive_pam_appa}

Given a total coded-bit budget
$R_{\mathsf{tot}}=\sum_{i=1}^{n-1}\log_2 M_i$, we adopt a variant of the APPA scheme in \cite{liu2025joint}, which directly minimizes the worst-case symbol error rate (SER) rather than maximizing mutual information as in Mercury/waterfilling (MWF) \cite{MWF_App2021}. The uncoded SER of the $i^{\text{th}}$ subchannel is
\begin{equation}
\mathsf{SER}_i =
2\left(1-{1}/{M_i}\right)
Q\!\left({\lambda_i p_i}/{\sigma_z}\right).
\end{equation}
With \(M_i=2^{m_i}\), \(m_i\in\mathbb{Z}_{\geq1}\), we impose \(m_i\geq1\) on all SVD modes in the high-SNR regime and optimize \(\{M_i,p_i\}_{i=1}^{n-1}\) as
\begin{align}
\min_{{M_i,p_i}}\quad&
\max_i \mathsf{SER}_i \notag\\
\mathrm{s.t.}\quad&
\sum_{i=1}^{n-1}\log_2 M_i=R_{\mathsf{tot}},\quad
\sum_{i=1}^{n-1}\frac{M_i^2-1}{3}p_i^2\leq nP.
\label{eq:joint_pam_min_ser}
\end{align}

For a target SER $\theta$ and a given $M_i$, the minimum required spacing coefficient is obtained exactly as
\begin{equation}
p_i=
\frac{\sigma_z}{\lambda_i}
Q^{-1}\!\left(
\frac{\theta}{2(1-1/M_i)}
\right).
\label{eq:optimal_pi_given_Mi}
\end{equation}
Since $M_i=2^{m_i}$, the resulting power cost for $m_i$ bits is
\begin{equation}
A_{m_i}(\theta)=
\frac{\sigma_z^2(2^{2m_i}-1)}{3\lambda_i^2}
\left[
Q^{-1}\!\left(
\frac{\theta}{2(1-2^{-m_i})}
\right)
\right]^2.
\label{eq:mode_cost_compact}
\end{equation}
The marginal cost of one additional bit is
\begin{equation}
\Delta_i(m_i;\theta)
=A_{m_i+1}(\theta)-A_{m_i}(\theta).
\label{eq:marginal_cost_compact}
\end{equation}
For the relevant PAM orders, $A_{m_i}(\theta)$ is discretely convex, so $\Delta_i(m_i;\theta)$ is non-decreasing in $m_i$. Thus, starting from $m_i=1$, the remaining $R_{\mathsf{tot}}-(n-1)$ bits are assigned greedily to the subchannel with the smallest marginal cost. A min-heap yields the optimal bit allocation for a given $\theta$, while one-dimensional bisection on $\theta$ determines the minimum achievable worst-case SER subject to \eqref{eq:joint_pam_min_ser}. The derivation and optimality proof are omitted due to space limitations.

\subsection{Exact Finite-Alphabet LLR Computation}
The parallel-channel representation \eqref{eq:dst_rx_s} also permits exact soft-output
demodulation. Recall that \(\mathcal X_{M_i}\) denotes the unit-spacing PAM-\(M_i\) alphabet in \eqref{eq:M_i}. Define
\begin{equation}
\mathcal X_{M_i}^{(k,b)}
=
\left\{
\ell\in\mathcal X_{M_i}:
\operatorname{bit}_k(\ell)=b
\right\},
\qquad b\in\{0,1\},
\label{eq:pam_bit_subsets}
\end{equation}
where $\operatorname{bit}_k(\ell)$ is the $k^{\text{th}}$ bit of the Gray label of
$\ell$, with \(k=1,\ldots,\log_2M_i\). The exact LLR for bit $k$ of
mode $i$ is
\begin{equation}
L_{i,k}
=
\log
\frac{
\displaystyle\sum_{\ell\in\mathcal X_{M_i}^{(k,0)}}
\exp\left(
-\frac{(r_i-\lambda_i p_i\ell)^2}{2\sigma_z^2}
\right)
}{
\displaystyle\sum_{\ell\in\mathcal X_{M_i}^{(k,1)}}
\exp\left(
-\frac{(r_i-\lambda_i p_i\ell)^2}{2\sigma_z^2}
\right)
}.
\label{eq:exact_pam_llr}
\end{equation}
Thus, $L_{i,k}>0$ favors bit $0$, while $L_{i,k}<0$ favors bit $1$. The computation is exact for finite-alphabet AWGN subchannels and can be stably implemented via log-sum-exp.

\vspace{-3pt}
\section{Simulation Results}
\label{sec:simulation}
\vspace{-2pt}
We compare the proposed SVD-based transceiver with TX-FFE/RX-DFE using three ISI-canceling taps ($t=3$) \cite{DFEFFE2020}, RX-MLSD ($t=0$) \cite{Forney1972}, and RX-SOVA ($t=0$) \cite{SOVA1989}. All schemes use a code rate of approximately $0.945$, consistent with prior work and high-speed Ethernet systems \cite{DFEFFE2020,ieee8023bs2017}. Fixed-PAM schemes employ PAM-$8$. We set $\alpha=0.8$ and $E_s=21$, and use Frame Error Rate (FER) as the primary metric. With the convention \(\sigma^2=N_0/2\) for real-valued AWGN and \(R_b\) denoting the average number of information bits per channel use, we set 

\begin{equation*}
\left({E_s}/{\sigma^2_z}\right)_{\mathrm{dB}}
=
\left({E_b}/{N_0}\right)_{\mathrm{dB}}
+3.0103
+10\log_{10}R_b.    
\end{equation*}

The first baseline uses $n=5{,}440$ coded PAM-$8$ symbols with a three-tap TX-FFE/RX-DFE, followed by three RS$(544,514)$ codewords over $\mathrm{GF}(2^{10})$~\cite{ieee8023bs2017} with symbol interleaving, yielding binary parameters $(16{,}320,15{,}420)$ and rate $0.945$. The second uses the same RS coding and symbol interleaving with RX-MLSD. The third uses $n=5{,}460$ coded PAM-$8$ symbols with RX-SOVA, followed by a regular LDPC code $(16{,}380,15{,}470)$ with bit interleaving, rate \(0.944\), variable-node degree \(d_v=3\), and check-node degree \(d_c=54\).

The proposed scheme uses $5460$ coded PAM symbols and one zero-padding channel use ($n=5,461$), with the same regular LDPC code $(16,380,15,470)$, bit interleaving, and adaptive PAM signaling.
The final channel input is fixed to zero to satisfy the single
zero-padding constraint. Thus, the proposed and SOVA systems use the same LDPC codeword length and information-bit payload; the proposed scheme incurs only the additional single zero-padding channel use.

\begin{figure}[tbp]
\centering
\includegraphics[width=0.45\textwidth]{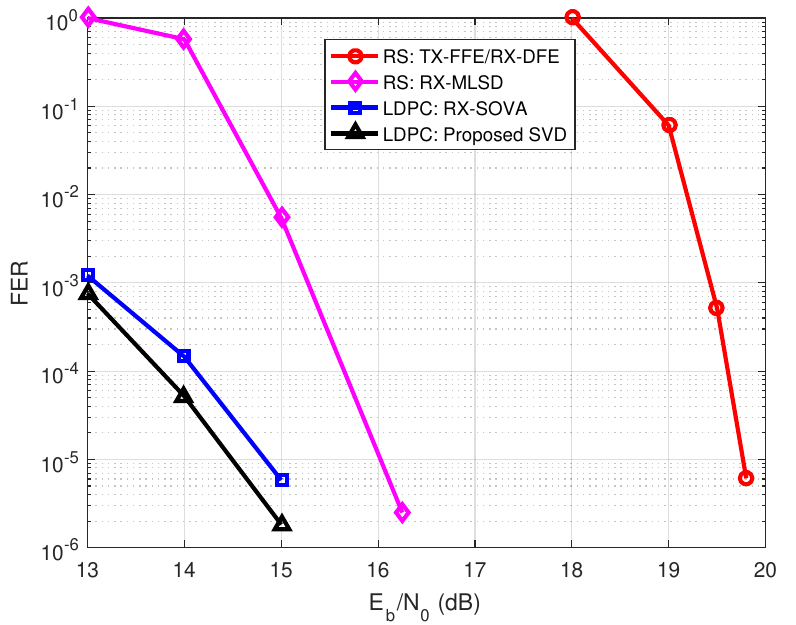} \vspace{-2mm} 
\caption{FER performance of different transceiver schemes over a memory-$1$ ISI channel with $\alpha=0.8$ and $E_s=P=21$.}
\label{fig:fer_comparison}
\vspace{-12mm}
\end{figure}

Figure~\ref{fig:fer_comparison} presents the resulting FER performance. At high SNR, the proposed SVD-based transceiver achieves comparable FER performance to the more computationally complex SOVA-based baseline. Both soft-decoding schemes substantially outperform
the RS-based hard-decoding architecture. These results
demonstrate the benefit of combining exact SVD diagonalization with
adaptive PAM signaling and soft LDPC decoding.

\section{Conclusion}
We presented a fast SVD-based transceiver for finite-length ISI channels.
A single zero-padding symbol exposes a structured channel matrix with a
closed-form SVD, whose singular-vector transforms are efficiently
implemented using FFTs. This enables exact SVD precoding
with $\mathcal{O}(n\log n)$ complexity and $\mathcal{O}(n)$ memory,
without explicitly forming the singular-vector matrices. The resulting
parallel AWGN modes further enable adaptive PAM, power
allocation, and exact LLR computation. Overall, the proposed single-zero-padding and FFT-based SVD framework provides a low-complexity foundation for large-PAM ISI transceivers in future high-throughput applications.

\noindent\textbf{AI Statement:}
Generative AI tools were not used to develop the methods, mathematical analysis, algorithms, simulations, or results presented in this paper.

\bibliographystyle{IEEEtran}
\bibliography{IEEEabrv,LIUBIB}

\end{document}